\documentclass[12pt]{article}
\usepackage{amsmath}
\usepackage{amsfonts}
\usepackage{amssymb}
\usepackage{latexsym}
\usepackage{graphicx}
\usepackage[english]{babel}
\input epsf.sty

\newcommand{\be}{\begin{equation}}
\newcommand{\ee}{\end{equation}}
\def\e{\tilde{\eta}}

\begin{document}

\title{
{\normalsize \hskip4.2in USTC-ICTS-07-15} \\{Generalized
Space-time Noncommutative Inflation}}
 \vspace{3mm}
\author{{Wei Xue$^{1}$, Bin Chen$^{1}$, Yi Wang$^{2,3}$}\\
{\small $^{1}$ Department of Physics, Peking University, Beijing 100871, P.R.China}\\
{\small $^{2}$ Institute of Theoretical Physics, Academia Sinica,
Beijing 100080, P.R.China}\\
{\small $^{3}$  The Interdisciplinary Center for Theoretical Study}\\
{\small of China (USTC), Hefei, Anhui 230027, P.R.China}}
\date{}
\maketitle

\begin{abstract}
  We study the noncommutative inflation with a time-dependent noncommutativity
between space and time. From the numerical analysis of power law
inflation, there are clues that the CMB spectrum
indicates a nonconstant noncommutative inflation. 
Then we extend our treatment to the inflation models with more
general noncommutativity  and find that the scalar perturbation
power spectrum depends sensitively on the time varying of the
spacetime noncommutativity. This stringy effect may be probed in
the future cosmological observations.
\end{abstract}

\section{Introduction}

As a successful theory describing the very early universe,
inflation\cite{inflation1,inflation2,inflation3} can solve many
problems in standard Big Bang cosmology, such as the large-scale
smoothness, horizon, and unexpected relic problems. More
importantly, inflation also predicts a scale invariant spectrum on
CMB which has been confirmed quite precisely in cosmological
experiments, such as WMAP three year result\cite{WMAP3yr}.
Although inflation has achieved many spectacular successes, it is
not perfect and has some conceptual problems\cite{Brand99}. One of
them is the trans-Plankian problem\cite{Martin:2000xs}. During
inflation, the physical wave length of each comoving mode expands
exponentially and escapes the horizon. The inflationary modes
which has just reentered the horizon at the present time
corresponds to a perturbation with a very small physical wave
length, perhaps smaller than the string scale or even the Planck
scale. So the usual assumption that these inflationary modes
originate from the local Minkowski vacuum is not robust. More
seriously, near the Planck scale, the quantum effects of gravity
should be large and
 the usual semi-classical treatment in the
inflationary models, in which the weakly self-coupled scalar field
theory is coupled to classical gravity, are known to break down.
An effective way to take the trans-Planckian effect into account
is to modify the dispersion relation of the
perturbations\cite{Martin:2000xs}. With the modified dispersion
relation, one can calculate the spectrum of the fluctuations and
compare with experiments. Therefore  the scale-invariant spectrum
in the inflationary cosmology gives a window to study the
trans-Planckian physics and hopefully has important implication
for a quantum gravity theory.

In \cite{Bran2002}, Brandenberger and Ho suggested that the
stringy spacetime uncertainty principle possibly affects the
primordial density fluctuations and leads to trans-Plankian
effects during inflation\footnote{For earlier attempts on relating
spacetime noncommutativity and inflationary cosmology, see Ref.
\cite{Chu:2000ww,Lizzi:2002ib}.}. Equivalently, one may realize
the uncertainty relation from the noncommutative relation \be
\label{nc2} [ t,x_p]=i {l_s}^2 \ ,\ee where $t$ and $x_p$ are
respectively physical time and physical radial coordinate. Here
choosing the radial coordinate makes sure that the
noncommutativity do not spoil the global symmetries of the whole
classical background. In practice, it is more convenient to
introduced another noncommutative relation
  \be \label{nc1}[ \tau,x]=i {l_s}^2,\ee
where $\tau$ is another time coordinate such that the
Friedmann-Robertson-Walker (FRW) metric for a spatially flat
universe could be written as \be \label{FRW}
ds^2=dt^2-a^2(t)dx^2=a^{-2}(\tau)d\tau^2-a^2(\tau)dx^2. \ee Note
that $\tau$ is not the conformal time and $x_p=a(t)x$. The scalar
inflation models has been studied by considering the scalar field
coupled to the classical background (\ref{FRW})  in terms of Moyal
star product induced by the noncommutative relation (\ref{nc1}). Due
to the noncommutativity, the coupling of the scalar field
fluctuations with the background is nonlocal in time. This leads to
a different spectral index from the usual commutative inflation in
the infrared region. And the model predicts sufficient running of
the spectral
index\cite{Bran2002,Huang:2003_1,Tsujikawa:2003gh,Huang:2003_2,Huang:2003fw}.
In recent researches, noncommutativity inflationary cosmology has
been generalized and extended in various ways: see Ref.
\cite{Liu:2004xg} for noncommutative tachyon inflation; see Ref.
\cite{Cai:2004ur} for curvature fluctuations of noncommutative
inflation; see Ref. \cite{huang:brane,Zhang:2006cc} for noncommutative brane
inflation; see Ref. \cite{Cai:2007et} for noncommutative eternal
inflation; the other related works could be found in
\cite{Kim:2004,Calcagni,Bamba:2004cu,Zhang:2006ta,Cai:2007xr}.

  It is believed that the trans-Plankian physics is closely related to
  the underlying fundamental theory of quantum gravity. In string theory,
  which is the best quantum gravity theory we have,  noncommutative geometry can
be realized quite naturally. One simple way is to consider the
Dp-brane in the presence of a constant Neveu-Schwarz Neveu-Schwarz
(NSNS) B-field in the flat spacetime background. The low energy
effective theory of Dp-brane is a noncommutative field
theory\cite{witten99}. In general, due to the presence of
nonconstant NSNS or Ramond-Ramond (RR) background and the curved
background, more general noncommutativity may appear on
Dp-brane\cite{nonassociate star product, Sethi,oh}. Though the
quantum field theory with the general noncommutativities has not
been studied as much as the usual case, they are important to the
study of the string theory. On the other hand, in some cosmology
scenario based on D-brane, like brane world scenario, the possible
noncommutativity on the brane could not be the one in (\ref{nc2})
due to the involved bulk flux and background. It is very possible
that the more precise cosmological observation data on CMB
spectrum in the future may shed more light on the trans-Planckian
physics, and even decide the details of the noncommutativity.
Therefore, it is interesting to investigate the physical
implications of the general noncommutativity in the inflationary
cosmology.
In this note we generalize the usual noncommutativity (\ref{nc2})
to the one with time dependence,  \be [ t,x_p]=i \theta (t)\
,\label{xtp} \ee and consider its cosmological implications.

The paper is organized as follows. In section 2 we first discuss the
time-dependent commutative relationship and introduce a new star
product with constant commutative relation in redefined space and
time coordinate. Then we analyze the power law inflation with a
time-dependent noncommutativity. Power law inflation exhibits
clearer effects of noncommutativity and our numerical analysis shows
that a time-dependent noncommutativity is quite reliably encoded in
the WMAP data. In section 3, we extend the discussion to the
inflation models with general scalar potential and general
noncommutativity. We develop a slightly new method to treat the slow
rolling noncommutative inflation and calculate the spectrum and the
spectral index. We find that the spectral index relies on the
time-varying of the noncommutative parameter very sensitively. We
conclude in Section 4.

\section{Power law noncommutative inflation}

  In the (quasi-)de Sitter spacetime, we assume that noncommutativity
of proper time and physical distance is $[ t,x_p]=i \theta (t)$. We
restrict that $\theta$ is only a function of time. This requirement
can preserve both the spatial translational and rotational symmetry
of the FRW metric. To describe the scalar field theory coupled with
the classical background, it is more convenient to define another
commutative relation
\be [\tau,x]_\ast=i l_s^2 \ee such that we can use the standard
Moyal product
 \be (f \ast g) (x,\tau)=\left.
e^{-{\frac{i}{2}}l_s^2 (\partial_{x}\partial{\tau'}-
\partial_{\tau}\partial_{y})}f(x,\tau)g(y,\tau')\right|_{y=x, \tau'=\tau}
 \ee
where x is the conformal coordinate, $x_p=a(t) x$, and $\tau$ is a
redefinition of the time variable, through the equation of $d\tau
=b(t)dt$. This leads to the commutative relation (\ref{xtp}) \be
[t,x_p]_\ast=i\frac{a}{b} {l_s}^2 =i\theta(t) \ . \ee Note that
$l_s$ is not the same as the string length, since $\sqrt{\theta}$ is
the effective string scale.
When $\theta$ is constant, the noncommutativity  returns to the
one in \cite{Bran2002}.

  In the power law inflation we have $a(t)=a_0
t^n$ and for simplicity we assume $b=a^{m/2}$, so the
time-dependent noncommutative parameter is  of the power law form,
 \be \theta(t)=a^{1-\frac{m}{2}} l_s^2.\ee
 Obviously the variable $m$ characterize the time-varying of the
 noncommutativity.
 The right hand side of
the equation may leak a constant coefficient, but it cannot affect
the final result by the redefinition of $a_0$. In this case,
$a(t)=a_0 t^n=\alpha_0\tau^{\frac{n}{m
n/2+1}}=(\tau/l)^{\frac{n}{m n/2+1}}$ , where $l$ is only the
constant coefficient. When we consider the more general inflation
potentials in the next section, all the power law assumptions,
including the assumption of $a(t)$ and $\theta (t)$, will
disappear.

  In the 1+1 dimensions, the noncommutative action reveals the fact
that gravitational background and scalar field interact with each
other nonlocally in time. The classical background is the flat FRW
metric
 \be ds^2 = -dt^2 + a(t)^2 dx^2 =
-b(\tau)^2 d{\tau}^2 + a(\tau)^2 dx^2.\ee
The action is
 \be S=\int d\tau dx \sqrt{-\det g}\frac{1}{2} \left(
\partial_{\tau}\phi^{\dagger}\ast b^{2}\ast\partial_{\tau}\phi
-(\partial_x\phi)^{\dagger}\ast a^{-2}\ast(\partial_x\phi) \right)
\ee Through Fourier transformation, the action in the momentum
space is \be S=V \int d\tau dk \frac{a}{2 b} ({\beta_k}^{+}
\partial_{\tau}\phi_{-k}\partial_{\tau}\phi_k
-{\beta_k}^{-} \phi_{-k}\phi_k ), \ee where \be
{\beta_k}^{+}(\tau)= \frac{1}{2}\left( b^{2}(\tau-l_s^2
k)+b^{2}(\tau+l_s^2 k) \right),\label{bp}\ee \be
{\beta_k}^{-}(\tau)= \frac{1}{2}\left( a^{-2}(\tau-l_s^2
k)+a^{-2}(\tau+l_s^2 k) \right).\label{bm}\ee One essential point
is that there exists a  cutoff on the wave number k in spatial
slices, due to the uncertainty principle of space and time \be
\Delta \tau \Delta x \geq l_s^2 \ ,\label{xtau}\ee which is
derived from the commutative relation \be [\tau, x]=l_s^2\ .\ee
According to the spacetime and energy-time uncertainty
relationship, we have a upper bound
 \be k \leq k_0(\tau)
\equiv
\left(\frac{{\beta_k}^{+}}{{\beta_k}^{-}}\right)^{1/4}{l_s}^{-1}\
. \label{initial condition}\ee It means that there is a
time-dependent minimal wavelength when we detect spacetime. Note
that the above bound is different from the one in \cite{Bran2002}
due to the time-dependence of the noncommutative parameter. This
will lead to different behavior in power spectrum, as we will show
soon.

  In inflationary scenario, the fluctuations in the UV and IR regions
have different spectral index. When the equal sign holds in
(\ref{initial condition}), we say that it gives the creating time
or the initial condition of $k$ modes. The UV region is the region
that the modes are created in the Hubble radius, while the IR
region is the region that the initial condition of modes are
already outside the Hubble radius. There is a turning point in
time that IR region converts to UV region. Modes, coinciding with
the initial condition, are created. When their wavelength is
Hubble radius
$k_0(\tau)=a H$, it is the time transforming between UV and IR.
 The turning point requests $\frac{k_0}{a}=H$. \be k_0=
\left(\frac{{\beta_k}^{+}}{{\beta_k}^{-}}\right)^{1/4}{l_s}^{-1}
=\{{\alpha_0}^{m+2}[(\tau+k l_s^2)^{\frac{m}{2}}(\tau-k
l_s^2)]^{\frac{2 n}{m n/2 +1}}\}^{1/4} l_s^{-1} =a H=a_0 t^{n+1}
\ee The larger $m$ we choose, the larger $k_0(\tau)$. If m is
becoming larger, the condition emerges later during inflation. It
also means that we can see the IR region in the sky earlier.

To calculate the power spectrum, we can write the action in the
effective conformal time coordinate
 \be S \simeq
V\int_{k<k_0} d\e dk \frac{1}{2} y_k^2(\e)
\left(\phi^{'}_{-k}\phi^{'}_k-k^2\phi_{-k}\phi_k\right)\ ,
 \ee where
 \be {d\e} =
 \left(\frac{{\beta_k}^{-}}{{\beta_k}^{+}}\right)^{1/2}d\tau=
 b\left(\frac{{\beta_k}^{-}}{{\beta_k}^{+}}\right)dt\equiv
 \frac{1}{\lambda a}dt\ ,\label{time}
\ee \be y_k = \frac{a}{b}({\beta_k}^{-} {\beta_k}^{+})^{1/4} \
.\label{y}\ee The primes denote derivatives with respect to $\e$.

  Considering the scalar perturbation in the $1+3$ dimensions, the action is
  \be S =
V\int_{k<k_0} d\e d^3k \frac{1}{2} z_k^2(\e)
\left(\phi^{'}_{-k}\phi^{'}_k-k^2\phi_{-k}\phi_k\right)\ ,
 \ee where $z_k$ is a smeared version of $z$:
  \be z_k=z y_k,\
z=\frac{a\dot{\phi}}{H}\ .\ee

  The scalar power spectrum can be calculated following \cite{MFB92}, and
be expressed as \be \mathcal{P}_{\mathcal{R}}=\frac{k^2}{4 \pi^2
z^2_k(\e_k)}\ .\label{Pk}\ee The choice of $\e_k$ is from \be
k^2=\frac{z{''}_k}{z_k}.\label{k}\ee

  First, consider the UV region. In the power law
inflation, $z\propto a$, up to a constant. According to the
relation between $d\tau$ and $d\eta$ (\ref{time}), we can derive
the relation between $\frac{d}{d\eta}$ and $\frac{d}{d\tau}$ in
the second order approximation. With (\ref{k}) we can get the time
when the modes exceeds the horizon, then using (\ref{Pk}), the
result of power spectrum is figured out. In the UV region, \be
\mathcal{P}_{\mathcal{R}}=\left(\frac{n(2n-1)}{(mn/2+1)}\right)^{\frac{n}{n-1}}\frac{n}{8
\pi^2}\left(\frac{l_p}{l}\right)^2l^{-\frac{2}{n-1}}k^{-\frac{2}{n-1}}\left(1+A\left(\frac{k_c}{k}\right)^{\frac
{m n-2 n-4}{n-1}}\right)\ , \label{power spectrum} \ee the
spectral index is \be
n_s=1+\frac{dln\mathcal{P}_{\mathcal{R}}}{dlnk}=1-\frac{2}{n-1}-\frac{mn/2-n+2}{n-1}A\left(\frac{k_c}{k}\right)^{\frac
{m n-2 n-4}{n-1}}\ ,\label{spectral index}\ee and the running of
spectral index is\be \alpha_s\equiv
\frac{dn_s}{dlnk}=\left(\frac{mn/2-n+2}{n-1}\right)^2A\left(\frac{k_c}{k}\right)^{\frac
{m n-2 n-4}{n-1}}\ ,\label{running of spectral index}\ee where
 \begin{eqnarray} A= \frac
{n}{8(n-1)(2n-1)(\frac{mn}{2}+1)^2}\left(\frac{m-4}{2}(m^3+m^2+14m+24)n^3 \right.\nonumber\\
\left. +(\frac{3}{2}m^3
+8m^2+22m+8)n^2+(-3m^2+22m+32)n+4(2-m)\right)\ ,\end{eqnarray} \be
k_c=l_s^{\frac{2n-2}{n}}l^{-\frac{m}{2}-1}\ ,\ee and $l_p$ is the
reduced plank length. When m=2, the results return back to the one
in \cite{Huang:2003_1}\cite{Huang:2003_2}. From the result,  we
have to require $(m-2)n\lesssim O(1)$ in order to have the scale
invariant spectrum.
If only considering the first order approximation, the result is
the same as the common inflation without noncommutativity.

  In the IR region, the effect of the noncommutativity is important.
  We can also take  second order
approximation as the UV region, but it is much more complicated
and is actually not necessary in the following discussions. So we
only give out the power spectrum in the first order approximation,
which turn out to be enough to illustrate the physical picture and
see the variation between different noncommutativities.

From the initial condition (\ref{initial condition}), we obtain
\be k\equiv
\left(\frac{{\beta_k}^{+}}{{\beta_k}^{-}}\right)^{1/4}{l_s}^{-1}
=\{{\alpha_0}^{m+2}[(\tau+k l_s^2)^{\frac{m}{2}}(\tau-k
l_s^2)]^{\frac{2 n}{m n/2 +1}}\}^{1/4} l_s^{-1}\ .\ee The equation
can be rewritten in the form \be(\frac{\tau}{k
l_s^2}+1)^{\frac{m}{2}}(\frac{\tau}{k l_s^2}-1) =
(\alpha_0^{m/2+1} l_s^2)^{-\frac{m n /2+1}{n}}(k
l_s^2)^{\frac{2}{n}+\frac{m}{2}-1}\ .\ee

  In the IR region, $k$ is so small that we can use the condition to simplify
the equation. Assume that $\tau \propto k^x$. Because the value of
the right hand side is not negative, $x\leq1$. If
$\frac{2}{n}+\frac{m}{2}-1>0$, the right hand side is a small
quantity; and if $\frac{2}{n}+\frac{m}{2}-1<0$, the right hand
side is a large quantity. When $x =1$, we can solve the equation
in the approximation of $\tau \simeq k l_s^2$, which requires $n
(1-\frac{m}{2})<2$. We obtain \be \tau=k l_s^2 +
(\frac{k^2l_s^2}{\alpha_0^{m/2+1}})^{\frac{m n /2+1}{n}}(2k
l_s^2)^{-\frac{m}{2}}\ .\ee Then,\be \mathcal{P}_{\mathcal{R}}\sim
k^2 z_k^{-2}(\e_k)\sim k^{\frac{\frac{3}{2} m n-3n+4}{\frac{1}{2}
m n +1}} \ .\ee If $x<1$, we can obtain \be x=\frac{m
n+2}{(\frac{m}{2}+1) n}, \ee and $\tau \sim k ^x$. From $x<1$, we
get $n (1-\frac{m}{2})>2$. Roughly speaking, this condition
requires $m<2$, i.e. the effective string length is decreasing
during inflation. Then, \be \mathcal{P}_{\mathcal{R}}\sim k^2
z_k^{-2}(\e_k)\sim k^0 \ .\ee The future cosmology evolution may
reveal the effect of IR region relating to the earlier period of
inflation. If $n (1-\frac{m}{2})<2$, the spectrum looks bluer than
the one in \cite{Bran2002}. If $n (1-\frac{m}{2})>2$, the spectrum
looks redder than \cite{Bran2002}. Furthermore the condition of $n
(1-\frac{m}{2})>2$ may reveal new phenomena that the spctrum is
scale invariant in the first approximation.

It would be illuminating to use our model to fix the experimental
data. There are four parameters, $l$, $l_s$, $m$, $n$, in our
model. From the WMAP data, we know \be
\mathcal{P}_{\mathcal{R}}=2.95\times 10^{-9} \quad \hbox{at}
  \quad k=0.002 \hbox{Mpc}^{-1} , \nonumber\ee
  \be n_s=0.93\pm 0.03, \quad  {dn_s\over d\ln k}=-0.031^{+0.016}
_{-0.017} \quad \hbox{at}\quad k=0.05 \hbox{Mpc}^{-1},\nonumber\ee
\be n_s=1.16\pm 0.10, \quad  {dn_s\over d\ln k}=-0.085\pm0.043
\quad \hbox{at}\quad k=0.002 \hbox{Mpc}^{-1}.\ee  Using
(\ref{power spectrum})(\ref{spectral index})(\ref{running of
spectral index}), we may determine the variation of
noncommutativity, i.e. the variable m. When we require the
$|A(\frac{k_c}{k})^{\frac {m n-2 n-4}{n-1}}|<0.5$, the numerical
analysis shows that the available data of m is 2.2 to 2.3. Since
$m>2$, it illustrates the fact that the noncommutativity is
decreasing during the inflation period.

\begin{figure}
\centering
\includegraphics[totalheight=2.5in]{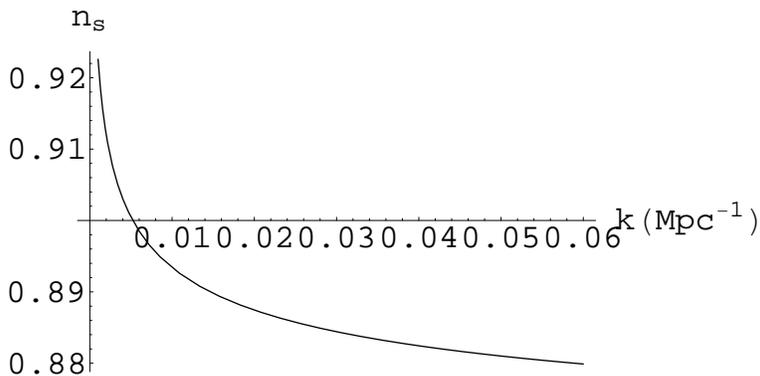}
\caption{\small{Here $n_s$ is the spectral index and k is the
comoving mode. As our observable universe gets larger in the
future, noncommutative inflation with decreasing noncommutativity
predicts that the spectrum will turn from red to blue clearly.}}
\label{fig:gr1}
\end{figure}

\begin{figure}
\centering
\includegraphics[totalheight=2.5in]{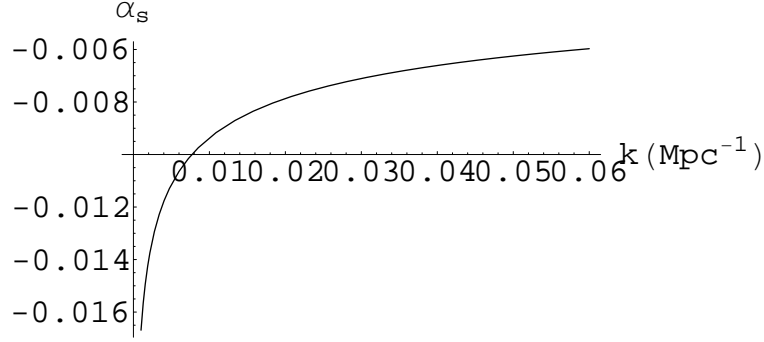}
\caption{\small{Here $\alpha$ is the running of the spectral index
and k is the comoving mode. The running becomes faster in the
later period. It is due to the stronger effects of
noncommutativity in the earlier period of inflation.}}
\label{fig:gr2}
\end{figure}

In the two figures above, we choose $n= 14.9$, $m= 2.2$, $l_s=
6.07 \times 10^{-32}\ {\rm cm}$, $l= 1.575 \times 10^{-23}\ {\rm
cm} $, which are in the allowable scope of the WMAP data. From the
numerical analysis above, we show that the WMAP data really
permits the existence of a time-dependent noncommutativity in the
power law inflation. Furthermore, our model predicts that the
spectrum will turn to bluer and the running of the index will be
faster.

\section{More general noncommutative inflation models}

In this section, we extend the above discussions on power law
inflation to more general cases, where a general inflaton
potential other than the exponential potential is considered.
Moreover, we also consider the more general spacetime
noncommutativity without the assumption $\theta(t)\sim
a^{1-\frac{m}{2}}$.
To simplify our discussion, here we assume a new slow roll
condition on the time dependent spacetime noncommutativity, namely
we require that :
\begin{equation}
  \sigma \equiv \frac{\dot{\theta}}{H\theta}, ~~~~~\left|\sigma\right|\ll 1.
\end{equation}

And as usual, we impose the slow roll conditions for  the Hubble
constant and the background inflaton field,
\begin{equation}
  \epsilon=-\frac{\dot{H}}{H^2},~~~\delta=-\frac{\ddot{\varphi}}{H\dot{\varphi}},~~~ |\epsilon|\ll 1, ~~~|\delta|\ll 1.
\end{equation}

The equation of motion for the perturbations can be written as
\cite{MFB92}
\begin{equation}
  \label{eq:eom}
  u_k{''}+\left(k^2-\frac{z_k{''}}{z_k}\right)u_k=0,
\end{equation}
where $u_k$ is the canonically normalized perturbation variable
$u_k\equiv -z_k \mathcal{R}_k$, and $\mathcal{R}_k$ is the
comoving curvature perturbation.

In the noncommutative case, ${z_k{''}}/{z_k}$ can be expanded as,
\begin{equation}
  \label{eq:z}
  \frac{z_k{''}}{z_k}=(\lambda^2a^2H^2)\left\{\frac{\ddot{z}_k}{H^2z_k}+\left(1+\frac{\dot{\lambda}}{H\lambda}\right)\frac{\dot{z}_k}{Hz_k} \right\},
\end{equation}
where as defined in (\ref{time}), $\lambda$ denotes the difference
between the usual conformal time and the modified conformal time
$\tilde\eta$.

With the  slow roll condition on $\theta$ in mind, we can expand
the corresponding quantities up to the leading order of the $l_s$
corrections as
\begin{equation}\label{expan}
  y_k=1+\mu,~~~\lambda=1-\mu,~~~\mu\equiv
  H^2\theta^2\frac{k^2}{a^2},
\end{equation}
The validity of this expansion requires $\mu<1$. This is the same
requirement as the perturbation is generated in the UV region.

Compared with the $\mu$ parameter defined in the work
\cite{Huang:2003fw}, $\mu\equiv H^2l_s^4{k^2}/{a^2}$, we see that
the time varying noncommutativity can be thought of as a
redefinition of $\mu$.

Using the slow roll approximation, we can integrate out the
modified conformal time as \footnote{In \cite{Huang:2003fw}, $\mu$
has been considered as nearly a constant. It
  is because the horizon crossing formula is used in
  \cite{Huang:2003fw} for calculating the power spectrum. So the time variable is set to the horizon
  crossing point. However in dealing with the equation of motion of a
  single comoving mode $k$, $\mu$
  should be varying very fast with time because $a$ expands exponentially.}
\begin{equation}\label{inttime}
\tilde{\eta}=\int\frac{dt}{a}(1+\mu)=-\frac{1}{aH}\left(1+\epsilon+\frac{1}{3}\mu\right).
\end{equation}

Insert (\ref{inttime}) and (\ref{expan}) into (\ref{eq:z}), the
equation (\ref{eq:eom}) becomes
\begin{equation}
  u{''}_k+ \left(1+\frac{8}{3}H^4\theta^2\right) k^2 u_k - \frac{2}{\tilde{\eta}^2}\left(1+3\epsilon-\frac{3}{2}\delta\right) u_k=0
\end{equation}
We see that in spite of using the modified conformal time, the
difference between this classical equation of motion and the usual
commutative one is only a rescale of $k$. As is shown in
\cite{Martin:2000xs,Chen:2006wn}, this rescale property also holds
in the quantum level for the initial conditions. So the comoving
curvature perturbation can be written as a function of
$\left(1+\frac{4}{3}H^4\theta^2\right) k$:
\begin{equation}
  \mathcal{R}_k=\mathcal{R} \left((1+\frac{4}{3}H^4\theta^2) k\right)
\end{equation}
And the perturbation spectrum takes the form
\begin{equation}
  \mathcal{P}_{\mathcal{R}}=\frac{k^3}{2\pi^2}\left|\mathcal{R} \left((1+\frac{4}{3}H^4\theta^2) k\right) \right|^2=\left(1-4H^4\theta^2\right)\frac{k^3\left|\mathcal{R}(k)\right|^2}{2\pi^2}
\end{equation}
where $\mathcal{R}(k)$ is the usual comoving curvature
perturbation calculated in the commutative inflation models.  We
have used the fact that $ \mathcal{P}_{\mathcal{R}}$ should
produce a nearly scale invariant spectrum. 

Then we have the spectral index $n_s$ and the running of the
spectral index $\alpha_s$,
\begin{equation}\label{nsm1}
  n_s-1\equiv \left.\frac{d\ln  \mathcal{P}_{\mathcal{R}}}{d\ln k}\right|_{k=aH}
  =16\epsilon\mu-4\epsilon+2\delta-8\sigma\mu \ ,\ \ \ \
  \alpha_s\equiv\frac{dn_s}{d\ln k}\ ,
\end{equation}
where $\mu$ is calculated at $k=aH$.

Note that it is the slow roll parameter $\sigma$ on $\theta$
rather than $\theta$ itself that appears in the spectral index.
Therefore even the noncommutativity is very tiny, its variation
could affect the spectral index strongly. From (\ref{nsm1}), it is
easy to find that if the noncommutativity $\theta(t)$ decreases
with time, the spectrum turns out to be bluer. While if
$\theta(t)$ increase with time, the spectrum turns out to be
redder.

Especially, if $\theta$ decreases rapidly at around 60 e-folds
before the end of inflation, then the power spectrum can be very
blue in the small multipole moment $l$ region. In this case, the
power spectrum can be greatly suppressed. This provides a possible
solution to the problem of the lack of the CMB anisotropy power on
the largest angular scales. Note that to make a very blue spectrum
with $n_s-1\sim {\cal O}(1)$, it is still possible that $\sigma$
and $\mu$ are both smaller than 1, so both the $\mu$ expansion and
the slow-roll approximation are still valid.

It is remarkable that in the discussion above, to treat the slow
rolling noncommutative inflation, we developed a slightly new
method to get the power spectrum and spectral index. To check its
validity, let us consider two special cases.

As the first example, let us consider the constant commutator
limit, where $\theta=l_s^2$. This has been considered in
\cite{Bran2002, Huang:2003fw}. In this case, the spectrum and the
spectral index agree completely with the ones in
\cite{Huang:2003fw}.

For another example, consider the power law inflation with a
monomial commutator. We have discussed this case in section 2,
using the method developed in \cite{Bran2002}. Once again, the
results of this section matches with the ones in Section 2
correctly.

The result (\ref{nsm1}) opens up a window for studying the
noncommutative inflation models. It shows that the spectral index
depends strongly on how $\theta$ evolves with time. Even a slow
rolling change of the noncommutativity could result in a very
different spectral index. This issue is interesting because, on
one hand, the question how the string theory
gives the trans-Planckian effect in the inflationary cosmology
should be taken into consideration seriously. For example, in some
 brane world scenario, the noncommutativity on the brane is
induced by the bulk background flux. In a curved time-dependent
background, how the flux evolves and give rise to the
time-dependent noncommutativity is an interesting question to ask.
There have already been some efforts to study string evolution in
the inflationary background \cite{Li:2007gf}. However a lot more
needs to be done to get a complete inflationary string theory
picture and determine the background field evolution during
inflation. On the other hand, it is exciting that precise
measurements on the CMB spectrum may provide a probe for the
underlying spacetime geometry emergent from string theory in a
dynamical way.

\section {Conclusion}

In this paper we discussed the noncommutative inflation models with
a time-dependent noncommutativity. In the power law inflation case,
we found that the WMAP data could encode the time dependence of the
noncommutativity. And our discussion on general inflationary models
showed that
the power spectral index sensitively relies on the time varying of
the noncommutativity. This suggests that the more precise
observatory data in the future may tell us the details of the
noncommutative inflationary scenario and probe the underlying
physics of quantum gravity.


\section*{Acknowledgments}

We would like to thank Bo-Qiang Ma ,Yi-Fu Cai, Xin Zhang for
dicussion. The work was partially supported by NSFC Grant No.
10405028,10535060, NKBRPC (No. 2006CB805905) and the Key Grant
Project of Chinese Ministry of Education (NO. 305001).

\end{document}